\documentclass[11pt]{article}
\usepackage{makeidx}
\usepackage{amssymb}
\usepackage{graphicx,vmargin,subfigure,float,amsmath,epsfig,epstopdf}
\usepackage{amsmath}

\input{tcilatex}
\begin{document}

\title{On Chaplygin Gas Braneworld Inflation with Monomial Potential}
\author{ A. Safsafi$^{1,2,3}$\thanks{%
Corresponding authors: safsafi\_adil@yahoo.fr, safsafiadil@gmail.com}{\small %
, } I. Khay$^{1,2}$, F. Salamate$^{1,2},$ H. Chakir$^{1,2,3}$ and M. Bennai$%
^{1,3}$ \\
$^{1}${\small \ Equipe Physique Quantique et Applications. }\\
$^{2}${\small Equipe de Recherche Subatomique et Applications.}\\
{\small Laboratoire de Physique de la Mati\`{e}re Condens\'{e}e , \ Facult%
\'{e} des Sciences\ Ben M'sik, }\\
\ {\small Universit\'{e} Hassan II, Casablanca, Maroc. }\\
{\small \ }$^{3}${\small Groupement National de Physique des Hautes
Energies, Focal point, LabUFR-PHE,\ Rabat, Maroc}}

\maketitle

\begin{abstract}
In this paper we study the Chaplygin gas model as a candidate for inflation
in the framework of the Randall Sundrum type-II braneworld model. We
consider the original and generalized Chaplygin gas model in the presence of
monomial potential. The inflationary spectrum perturbation parameters are
reformulated and evaluated in the high-energy limit and we found that they
depend on several parameters. We also showed that these perturbation
parameters are widely compatible with the recent Planck data for a
particular choice of the parameters space of the model. A suitable
observational central value of $n_{s}\simeq $ $0.965$ is also obtained in
the case of original and generalized Chaplygin gas.


\end{abstract}

\date{}
\tableofcontents

\section{Introduction}

It is widely believed that the early universe underwent a period of
accelerated expansion called inflation \cite{AH}, which has become the
standard paradigm of modern cosmology. Inflation model has been proposed as
an attempt to solve the shortcoming of the standard Big-Bang model of
cosmology, known usually as the flatness and the horizon problem \cite{AD}%
. Inflation is the candidate for understanding the physics of the very early
universe, typing the evolution of the universe to the properties of one or
more scalar inflaton fields, responsible for creating an accelerating
expanding universe \cite{Liddle}. Recently, a great amount of work has been
invested in studying the inflationary model with several candidate. Among
these candidates, the Chaplygin gas model \cite{DN}. This model was also
used to describe an mysterious dark sector so-called dark energy and dark
matter \cite{DM} and also used to describe the early universe 
\cite{Bouhmadi}. In cosmology, dark energy is a form of unknown energy
dominating the universe with a hugely negative pressure \cite{OLuongo}. It
is demonstrated by various astrophysical observations, including the
accelerating universe \cite{LXu}. Recall that the dark energy is defined
by an exotic equation of state of the form $P_{DE}=\omega \rho _{DE},$ where 
$P_{DE}$ and $\rho _{DE}$ are the pressure and energy density of dark
energy, while $\omega $ is the equation of state parameter of the dark
energy \cite{MubasherJamil}. There are several candidates to describe the
dark matter and dark energy in the cosmology \cite{JGleyzes}. Among them
K-essence model \cite{EGuendelman} , the $\Lambda $CDM model \cite{Orlando}%
, cosmological constant \cite{MBronstein}, tachyon model \cite{MRSetare} and quintessence model \cite{PFGonzalez}. There are
different kinds of the Chaplygin gas model which have been proposed in the
literature. For example\ the general model named extended Chaplygin gas
which is defined by an exotic equation of state of the form \cite{EOKahya}%
\begin{equation*}
\ p=\sum B_{m}\rho ^{m}-\frac{A}{^{\rho ^{\alpha }}}
\end{equation*}%
where $B_{m}$, $A$, $m$ are universal positive constants, and $0\prec \alpha
\preceq 1$. In the recent years the extended Chaplygin gas was the subject
of several cosmological and phenomenological studies \cite{FRW}. Note
that, when $m=1$ we obtain the case of modified Chaplygin gas \cite{Benaoum}$.$ The original Chaplygin gas corresponds to the case 
$B_{m}=0$ and $\alpha =1,$ this type was studied in \cite{RZarrouki}$.$ In the case of $B_{m}=0$ and $\alpha \neq 1$ it is known as the generalized Chaplygin gas (GCG) \cite{JRVillanueva} and finally for $B_{m}=0$ and $\alpha =0$ corresponds to the $\Lambda $CDM model \cite{Orlando}. Further, the Chaplygin gas models has been studied in different paper. In the Ref. \cite{SBNassur} the authors have studied the interaction of the dark energy with some fluids specially Chaplygin gas in the context of $f(T)$ theory$.$ In \cite{SouravDutta}, the authors have proposed an attempt for emergent universe scenario with modified Chaplygin gas, and have shown that it is not possible to have emergent scenario with model. In another work  \cite{BPourhassan}, B. Pourhassan et all have examined extended model of Chaplygin gas equation of state for which it recovers barotropic fluid with quadratic equation of state, and have found that extended Chaplygin gas may be a more appropriate model than generalized and modified Chaplygin gas and give a best fit with the observational data.

On the other hand, the Chaplygin gas inspired inflation model \cite{OBertolami} was the subject of several cosmological and phenomenological studies. In this context, the scalar field is usually the standard inflaton field, where the energy density can be extrapolated to obtain a successful inflationary period with a Chaplygin gas model. In the same context, a work has been done by R. Herrera \cite{Herrera1} where the brane-Chaplygin inflationary model was studied in great details and considered as a viable alternative model that can provide an accelerated expansion of the early universe. In extension of the Ref. \cite{Herrera1}, the similar work was performed for the case of the tachyon-Chaplygin inflationary model by using an exponential potential in the high-energy regime \cite{Herrera2}.

In the present paper, we are going to use the Randall-Sundrum II braneworld
model to study the original and generalized Chaplygin gas as a candidate for
the primordial inflation by assuming that the matter source on the brane
consist of a Chaplygin gas. The Chaplygin gas emerges as a effective fluid
of a generalized d-brane in a ($d+1$, $1$) spacetime, where the action can
be written as a generalized Born--Infeld action \cite{Bento}. These models
have been extensively studied in the literature \cite{Bouhmadi}. The
motivation for introducing Chaplygin-brane scenarios is the increasing
interest in higher-dimensional cosmological models, motivated by superstring
theory, where the matter fields are confined to a lower-dimensional brane
while gravity can propagate in the bulk. On the other hand, the Chaplygin
gas model seems to be a viable alternative to models that provide an
accelerated expansion of the early universe. Our aim is to quantify the
modifications of the Chaplygin inspired inflation in the Braneworld scenario.%
\textbf{\ }We use the monomial potential to study various perturbation
spectrum parameters such as the scalar spectral index $n_{s}$ and the ratio $%
r$ and the running of the scalar spectral index $\frac{dn_{s}}{d\ln (k)}$ in
the high-energy limit, particularly for a suitable choice of the different
parameters. We show that the inflation parameters are in good agreement with
recent Planck 2015 data \cite{Planck2015}.

An outline of the remainder of this paper is as follows: We first begin in
section 2, by recalling the standard inflation and Chaplygin gas Braneworld
inflation formalism, in particular the modified Friedmann equation. In
section 3, we study different perturbation spectrum concerning monomial
potential in the High-energy limit, and we present our results for original
and genaralized Chaplygin gas on the brane. The last section is devoted to
conclusion.

\section{Genaralized Chaplygin gas braneworld inflation}

\subsection{\protect\bigskip\ Inflationary Universe}

In this section we propose a short description on standard inflation
formalism to enable readers to understand the meanings of terms here
involved. Inflation is the period of the early universe that undergoes an
accelerating phase \cite{AH}. This period is equivalent to $\ddot{a}$ $%
\succ 0.$ The candidate that can yield this acceleration phase and
responsible for driving inflation is a scalar field named inflaton field.
The pressure and energy of the inflaton field are given by $p_{\phi }=\frac{1%
}{2}\dot{\phi}^{2}-V(\phi )$ and $\rho _{\phi }=\frac{1}{2}\dot{\phi}%
^{2}+V(\phi )$, where $V(\phi )$ is the scalar potential. The Friedmann
equation reads $H^{2}=\frac{8\pi G}{3}\left[ \frac{1}{2}\dot{\phi}%
^{2}+V(\phi )\right] .$ During inflation, the potential $V(\phi )=V$ depends
only on the inflaton field $\phi $. It is supposed that the field equation $%
\ddot{\phi}+3H\dot{\phi}=-V^{\prime }(\phi )$, is well-approximated by $3H%
\dot{\phi}=-V^{\prime }(\phi ).$ The condition for acceleration requires
that $\dot{\phi}^{2}\prec V,$ This is called the slow-roll approximation.
With very small $\dot{\phi}^{2}$, the Friedmann equation is approximately $%
3M_{p}^{2}H^{2}\simeq V,$~and the flatness conditions $\varepsilon $ $\prec
\prec 1$ and $\mid \eta \mid \prec \prec 1,$ where $\varepsilon =\frac{1}{2}%
M_{p}^{2}\left( \frac{V^{\prime }}{V}\right) ^{2}$ and $\eta =M_{p}^{2}\frac{%
V^{\prime \prime }}{V^{\prime \prime }}.$ The amount of inflation can be
measured in term of the e-folding number $N$ given by equation $%
N=M_{p}^{-2}\int_{\phi _{end}}^{\phi _{\ast }}\left( \frac{V}{V^{\prime }}%
\right) d\phi ,$ where $\phi _{\ast }$ and $\phi _{end}$ are the values of
the scalar field at the epoch when the cosmological scales exit the horizon
and at the end of inflation, respectively.

The small quantum fluctuations in the scalar field lead to fluctuations in
the energy density which was studied in a perturbative theory \cite{David}.
As discussed in \cite{Maartens} quantum fluctuations effect of the inflaton
are generally negligibles, since the coupling of the scalar field to bulk
gravitational fluctuations only modifies the usual 4D predictions at the
next order in the slow-roll expansion. So, one can define the power spectrum
of the curvature perturbations as $P_{R}\left( k\right) =\left( \frac{H^{2}}{%
2\pi \overset{\cdot }{\phi }}\right) ^{2}$. On the other hand, the quantum
fluctuations in the scalar field lead also to fluctuations in the metric 
\cite{Langlois}. In this way, one can define the amplitude of tensor
perturbations as $P_{g}\left( k\right) =\frac{8}{M_{p}^{2}}\left( \frac{H}{%
2\pi }\right) ^{2}F^{2}\left( x\right) .$ These results lead to the ratio of
tensor to scalar perturbations $r=\frac{P_{g}\left( k\right) }{P_{R}\left(
k\right) }.$ In relation to $P_{R}\left( k\right) $, the scalar spectral
index is defined as $n_{s}-1=\frac{d\ln P_{R}\left( k\right) }{d\ln \left(
k\right) }.$ Refs. \cite{Liddle} are recommended for further reading on
inflation.

\subsection{Genaralized Chaplygin gas on the brane}

Braneworld inflation \cite{Braneworld} is a particular kind of inflation
models. It is based primarily on the cosmological model Randall-Sundrum type
II\ which describes the universe in five dimensions with the presence of a
brane that includes all ordinary matter. The generalized Chaplygin gas is a
perfect fluid characterised by the following equation of state \cite{PingXi}%
: 
\begin{equation}
p=-\frac{A}{\rho ^{\alpha }}  \label{1}
\end{equation}%
where $\rho $ and $p$ are the energy density and pressure of the generalized
Chaplygin gas, respectively, $\alpha $ is a constant satisfaying $0\prec
\alpha \preceq 1$, and $A$ is a positive constant.

Inserting the equation (\ref{1}) in the equation of conservation of energy $%
\dot{\rho}+3H(\rho +p)=0,$ we obtain the following expression for the energy
density

\begin{equation}
\rho _{ch}=\left[ A+\left( \rho _{ch0}^{\alpha +1}-A\right) \left( \frac{%
a_{0}}{a}\right) ^{3(\alpha +1)}\right] ^{\frac{1}{\alpha +1}}  \label{2}
\end{equation}
where $a_{0}$ and $\rho _{ch0}$\ are the current values of the scale factor
and the generalized Chaplygin gas energy density, respectively.

The modification of the equation (\ref{4}) is realized from an extrapolation
of equation (\ref{2}), where the density matter $\rho _{m}\thicksim a^{-3}$
is replaced by the scalar field as $\rho =\left[ A+\rho _{m}^{\left( \alpha
+1\right) }\right] ^{\frac{1}{\alpha +1}}\rightarrow \rho =\left[ A+\rho
_{\phi }^{\left( \alpha +1\right) }\right] ^{\frac{1}{\alpha +1}}.$

In this section, we will recall briefly some basic facts of Randall-Sundrum
type II braneworld model \cite{RSII1}. We suppose that the univers is
filled with a perfect fluid with energy density $\rho (t)$ and pressure $%
p(t) $ in which the Friedmann equation is modified from its usual form \cite{Herrera1}

\begin{equation}
H^{2}=k\rho _{\phi }\left[ 1+\frac{\rho _{\phi }}{2\lambda }\right] +\frac{%
\Lambda _{4}}{3}+\frac{\xi }{a^{4}}
\end{equation}%
where $H=\frac{\dot{a}}{a}$ defines the Hubble parameter, $\rho _{\phi }$
represents the matter confined to the brane, $k=\frac{8\pi G}{3}=\frac{8\pi 
}{3M_{p}^{2}}$, $\Lambda _{4}$ the current cosmological constant $\xi $ is
an integration constant and thus transmitting bulk graviton influence onto
the brane. This term appears as a form of \textquotedblleft dark
radiation\textquotedblright\ and may be fixed by observation \cite{P}.
However, during inflation this term is rapidly diluted, so we will neglect
it. Where $\lambda $ is the brane tension, $M_{p}$ is the four-dimensional
planck mass, which is related to the five-dimensional $M_{5}$ by $M_{p}=%
\sqrt{\frac{3\text{ }M_{5}^{6}}{4\pi \lambda }}$. Note that the crucial
correction to standard inflation is given by the density quadratic term $%
\rho ^{\left( \alpha +1\right) }.$ Note also that in the limit $\lambda $ $%
\rightarrow $ $\infty $, we recover standard four-dimensional general
relativistic results.

The Friedmann equation will become \cite{Herrera2}

\begin{equation}
H^{2}=\frac{8\pi }{3M_{p}^{2}}(A+\rho _{\phi }^{\left( \alpha +1\right) })^{%
\frac{1}{\alpha +1}}\left[ 1+\frac{(A+\rho _{\phi }^{\left( \alpha +1\right)
})^{\frac{1}{\alpha +1}}}{2\lambda }\right] ,  \label{4}
\end{equation}

In four-dimensional general relativity, the condition for inflation is $\dot{%
\phi}^{2}\prec \prec V(\phi ),$ i.e\textbf{\ }$p_{\phi }\prec -\frac{1}{3}%
\rho _{\phi },$ where $p_{\phi }=\frac{1}{2}\dot{\phi}^{2}-V(\phi )$ and $%
\rho _{\phi }=\frac{1}{2}\dot{\phi}^{2}+V(\phi )$, $V(\phi )=V$ is the
scalar potential and $\phi $ is the inflaton field the scalar field
satisfies the Klein-Gordon equation :

\begin{equation}
\ddot{\phi}+3H\dot{\phi}+V^{\prime }(\phi )=0.
\end{equation}

Note that $\dot{\phi}=\frac{\partial \phi }{\partial t},$ $\ddot{\phi}=\frac{%
\partial ^{2}\phi }{\partial t^{2}},$ $V^{\prime }=\frac{\partial V}{%
\partial \phi }.$ During inflation, the relation between the energy density
and the scalar potential is $\rho _{\phi }\simeq V$, we consider the
slow-roll approximation $\ \dot{\phi}^{2}\prec \prec V(\phi )$ and $\ddot{%
\phi}\prec \prec 3H\dot{\phi}$ the Friedmann equation reduces to:

\begin{equation}
H^{2}=\frac{8\pi }{3M_{p}^{2}}(A+V^{\left( \alpha +1\right) })^{\frac{1}{%
\alpha +1}}[1+\frac{(A+V^{\left( \alpha +1\right) })^{\frac{1}{\alpha +1}}}{%
2\lambda }].  \label{6}
\end{equation}

The term in square brackets is the brane-modification to the standard
slow-roll expression for the Hubble rate.

We consider the slow-roll parameters, in the Randall-Sundrum type II
Braneworld model \cite{Maartens}. The two first parameters are given for
generalized Chaplygin gas model by :

\begin{equation}
\varepsilon =\frac{M_{P}^{2}}{16\pi }\frac{V^{\alpha }V^{^{\prime 2}}}{%
(A+V^{(\alpha +1)})^{\frac{\alpha +2}{\alpha +1}}}\left[ \frac{1+\frac{%
(A+V^{\left( \alpha +1\right) })^{\frac{1}{\alpha +1}}}{\lambda }}{%
^{^{\left( 1+\frac{(A+V^{\left( \alpha +1\right) })^{\frac{1}{\alpha +1}}}{%
2\lambda }\right) ^{2}}}}\right] ,
\end{equation}

and

\begin{equation}
\eta =\frac{M_{P}^{2}}{8\pi }\frac{V^{\prime \prime }}{(A+V^{(\alpha +1)})^{%
\frac{1}{\alpha +1}}}\left[ \frac{1}{1+\frac{(A+V^{\left( \alpha +1\right)
})^{\frac{1}{\alpha +1}}}{2\lambda }}\right] .
\end{equation}

The inflationary phase ends when $\varepsilon $ or $\mid \eta \mid $are
equal to one, during inflation, the conditions $\varepsilon $ $\prec \prec 1$
and $\mid \eta \mid \prec \prec 1$ are satisfied. Note that, At low
energies, $V$ $\ll $ $\lambda $, the slow-roll parameters reduce to the
standard inflation.

In addition, the number of e-folding is given by

\begin{equation}
N=\frac{-8\pi }{M_{P}^{2}}\int_{V_{\ast }}^{V_{end}}\frac{\left(
A+V^{(\alpha +1)}\right) ^{\frac{1}{\alpha +1}}}{V^{\prime 2}}[1+\frac{%
(A+V^{\left( \alpha +1\right) })^{\frac{1}{\alpha +1}}}{2\lambda }]dV,
\label{NN}
\end{equation}

where $V_{\ast }$ and $V_{end}$ are the values of the potentials at the
horizon exit and the end of inflation, respectively.

The inflationary spectrum perturbation is produced by quantum fluctuations
of fields around their homogeneous background values. The small quantum
fluctuations in the scalar field lead to fluctuations in the energy density
and in the metric, for that, we define the power spectrum of the curvature
perturbations by \cite{Maartens} :

\begin{equation}
P_{r}(k)=\left( \frac{H^{2}}{2\pi \dot{\phi}}\right) ^{2}  \label{10}
\end{equation}

by using the equation (\ref{6}) and (\ref{10}), we find the expression for
power spectrum of the curvature perturbations :

\begin{equation}
P_{r}(k)=\frac{128\pi }{3M_{p}^{6}}\frac{(A+V^{\left( \alpha +1\right) })^{%
\frac{3}{\alpha +1}}}{V^{^{\prime }2}}\left[ 1+\frac{(A+V^{\left( \alpha
+1\right) })^{\frac{1}{\alpha +1}}}{2\lambda }\right] ^{3}.
\end{equation}

Another important inflationary spectrum parameter is the amplitude of the
tensorial perturbations $Pg(k)$, describing the primordial gravitational
wave perturbations produced by a period of extreme slow-roll inflation,
which is defined by \cite{Langlois}:

\begin{equation}
P_{g}=\frac{64\pi }{M_{p}^{2}}\left( \frac{H}{2\pi }\right) ^{2}F^{2}(x),
\end{equation}

where $x=HM_{p}\sqrt{\frac{3}{4\pi \lambda }\text{ }}$ and $F^{2}(x)=\left( 
\sqrt{1+x^{2}}-x^{2}\sinh ^{-1}(\frac{1}{x})\right) ^{-1}.$ Note that in the
low-energy limit $(A+V^{\left( \alpha +1\right) })^{\frac{1}{\alpha +1}%
}\prec \prec \lambda ,$ we have $F^{2}(x)\simeq 1,$ and in the high-energy
limit $(A+V^{\left( \alpha +1\right) })^{\frac{1}{\alpha +1}}\gg \lambda ,$ $%
F^{2}(x)\simeq \frac{3}{2}x=\frac{3}{2}\frac{(A+V^{\left( \alpha +1\right)
})^{\frac{1}{\alpha +1}}}{\lambda }$.

We define the ratio $r$ of tensor to scalar as :

\begin{equation}
r=\left( \frac{P_{g}(k)}{P_{r}(k)}\right) _{k=k^{\ast }}.
\end{equation}

Here $k^{\ast }$ correspond to the case $k=Ha,$ the value when the universe
scale crosses the Hubble horizon during inflation. From equations (11, 12),
the tensor to scalar is giving by :

\begin{equation}
r=\frac{M_{p}^{2}}{\pi }\frac{V^{\prime 2}F^{2}(x)}{(A+V^{\left( \alpha
+1\right) })^{\frac{2}{\alpha +1}}\left( 1+\frac{(A+V^{\left( \alpha
+1\right) })^{\frac{1}{\alpha +1}}}{2\lambda }\right) ^{2}}.
\end{equation}

The scalar spectral index is presented by \cite{David}:

\begin{equation}
\begin{array}{llll}
n_{s}-1 & = & \frac{d\ln P_{R}(k)}{d\ln (k)} &  \\ 
& = & \frac{M_{p}^{2}}{8\pi (A+V^{\left( \alpha +1\right) })^{\frac{1}{%
\alpha +1}}\left( 1+\frac{(A+V^{\left( \alpha +1\right) })^{\frac{1}{\alpha
+1}}}{2\lambda }\right) }\left( -3\frac{V^{\alpha }V^{\prime 2}}{\left(
A+V^{\alpha +1}\right) }\frac{\left( 1+\frac{(A+V^{\left( \alpha +1\right)
})^{\frac{1}{\alpha +1}}}{\lambda }\right) }{\left( 1+\frac{(A+V^{\left(
\alpha +1\right) })^{\frac{1}{\alpha +1}}}{2\lambda }\right) }+2V^{\prime
\prime }\right) . & 
\end{array}%
\end{equation}

The running of the scalar index is also defined as :

\begin{equation}
\frac{dn_{s}}{d\ln (k)}=\frac{M_{p}^{2}}{4\pi }\frac{V^{\prime }}{
(A+V^{\left( \alpha +1\right) })^{\frac{1}{\alpha +1}}}\frac{1}{(1+\frac{
(A+V^{\left( \alpha +1\right) })^{\frac{1}{\alpha +1}}}{2\lambda })}\left( 3 
\frac{\partial \varepsilon }{\partial \phi }-\frac{\partial \eta }{\partial
\phi }\right) .
\end{equation}

Note that in the limit $A\rightarrow 0$, the perturbation spectrum
parameters coincides with brane-inflation \cite{Maartens} in particular for $%
\alpha =1$. Also, in the low-energy limit, $(A+V^{\left( \alpha +1\right)
})^{\frac{1}{\alpha +1}}\ll \lambda $, the slow-parameters reduce to the
standard form \cite{OBertolami}.

In what follows, we shall apply the above Braneworld formalism with a
monomial potential in the high-energy limit; i.e. $(A+V^{\left( \alpha
+1\right) })^{\frac{1}{\alpha +1}}\gg \lambda ,$ in relation to recent
Planck data.

\section{Chaplygin gas with monomial potential}

\subsection{Original Chaplygin gas}

In the following, we will concentrate on the original Chaplygin gas, it is
reached as a special case of the general Chaplygin gas, it is proposed as
possible explanations of the acceleration of the current univers. The
original Chaplygin gas model has been extensively studied. For exemple, in 
\cite{JCFabris} study the behaviour of density perturbations in an
Universe dominate by the Chaplygin gas, and found that in spite of
presenting negative pressure of Chaplygin gas, is stable at small scale,
which opposite to in general what happens with perfect fluids with negative
pressure. In \cite{RZarrouki}, the authors have focused to study a
Chaplygin gas model in Braneworld inflation with an exponential potential,
and have obtained for negligible and small running of the scalar spectral
index, the inflationary parameters are in good agreement with observation
data. Another example \cite{HongshengZhang}, the authors have analysed a
phase space of the evolution for a Friedmann Robertson Walker universe
driven by an interacting of Chaplygin gas and dark matter, their results are
derived from continuity equations, which means that they are independent of
any theories of gravity. The original Chaplygin gas model is characterized
by an exotic equation of state of the form 
\begin{equation}
p=-\frac{A}{\rho },
\end{equation}%
where $A$ is a positive constant.

In this section we will propose to investigate monomial potential in
braneworld context with generalized Chaplygin gas. This potential used in
very recent model in diffirent work. In the paper \cite{MayraJReyes} the authors study the attractors solutions of the dynamical
system of a scalar field endowed with monomial potentials, and shown that
the behaviour found for monomial potentials is typical in realistic
inflationary models. Furthermore, R.Zarrouki et all have\emph{\ }studied
various inflationary spectrum perturbation parameters with three types of
potentials they shown that the monomial potential provides the best fit
results to observations data \cite{Zarrouki}. This potential is given by%
\begin{equation}
V=M\phi ^{n},
\end{equation}%
where $n$ is an positive intege and $M$ is a parameter of dimension $%
[E]^{4-n}$. In order to derive the inflationary parameters $n_{s},$ $r$ and $%
\frac{dn_{s}}{d\ln (k)}$, let us consider the monomial potential . In this
case, the scalar spectral index $n_{s}$, the ratio $r$ and the running of
the scalar index $\frac{dn_{s}}{d\ln (k)}$ becomes:%
\begin{equation}
n_{s}=\frac{M_{p}^{2}\lambda }{2\pi \left( A+M^{2}\phi _{\ast }^{2n}\right) }%
\left( -\frac{3n^{2}M^{3}\phi _{\ast }^{3n-2}}{\left( A+M^{2}\phi _{\ast
}^{2n}\right) }+n\left( n-1\right) M\phi _{\ast }^{n-2}\right) +1,
\end{equation}

\begin{equation}
r=\frac{6M_{p}^{2}\lambda M^{2}n^{2}\phi _{\ast }^{2n-2}}{\pi \left(
A+M^{2}\phi _{\ast }^{2n}\right) ^{\frac{3}{2}}},
\end{equation}%
and%
\begin{equation}
\frac{dn_{s}}{d\ln (k)}=-\frac{M_{p}^{4}\lambda ^{2}n^{2}(2n+1)(n+2)}{8\pi
^{2}\phi _{\ast }^{4}(A+(M\phi ^{n})^{2})}
\end{equation}%
Note that in the limit $A\rightarrow 0$, the scalar spectral index $n_{s}$,
the ratio $r$ and $\frac{dn_{s}}{d\ln (k)}$ coincides with Ref \cite{Zarrouki}.

Although we have analytic results for slow-roll parameters and the number of
e-folding, it is not easy to solve them to obtain $\phi _{\ast }$ at which
the observables $n_{s}$, $r$ and $\frac{dn_{s}}{d\ln (k)}$ should be
evaluated. For it, we proceed numerically by finding $\phi _{end}$ and using
the Eq (\ref{NN}) to obtain $\phi _{\ast }$ while making sure that the
slow-roll parameters remain small in this range of $\phi $.

To complete our study with original chaplygin gas, we analyse the variations
of the perturbation spectrum parameters with respect to $N$ for various
values of $n=1,$ $2,$ $3,$ $4.$ We can see that these observables depends on
several parameters$.$ For\ this\ purpose,\ we\ take\ the\ inflationary\ scale%
\emph{\ }$M\sim O(10^{15}$ $GeV),$ the brane tension value $\lambda \sim O$ (%
$10^{68}GeV^{4})$ and $A\sim 10^{-13}M_{p}^{8}$ \cite{Herrera1}$,$ in order
to obtain consistent perturbation spectrum parameters with recent Planck
data.

\begin{figure}[tbp]
\centering\includegraphics[width=0.7\linewidth]{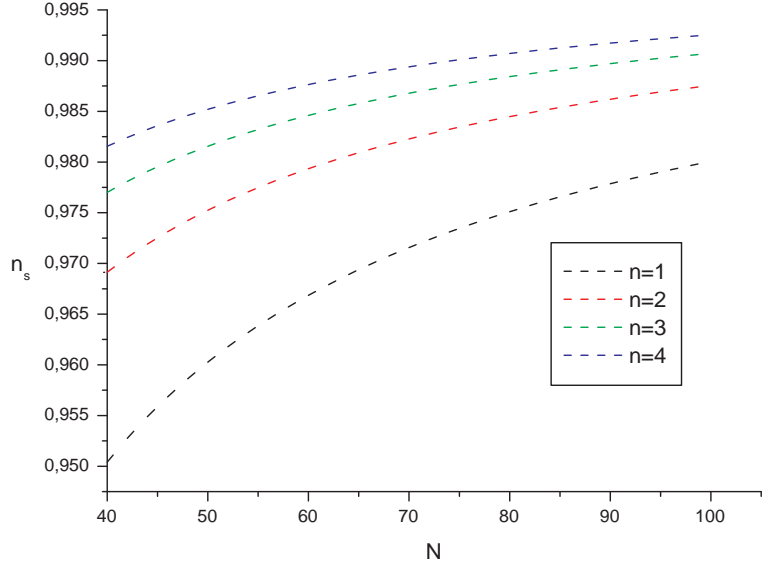}
\caption{$n_{s}$ versus $N$ for various values of $n.$ We take $M\sim
O(10^{15}$ $GeV),$ $\protect\lambda \sim O$ ($10^{68}GeV^{4})$ and $A\sim
10^{-13}M_{p}^{8}.$}
\label{fig1}
\end{figure}

Figure \ref{fig1} present the variations of scalar spectral index $n_{s}$ as
a function of e-folding number $N$. The scalar spectral index $n_{s}$ have a
increasing behaviour as we increase the e-folding number $N$. We also note
that the values of $n_{s}$ are found to be consistent with the Planck data
for large domain of $N$ and the central value of $n_{s}\simeq $ $0.965$ is
obtained in particular for $n=1;2$.

\begin{figure}[]
\centering
\includegraphics[width=0.7\linewidth]{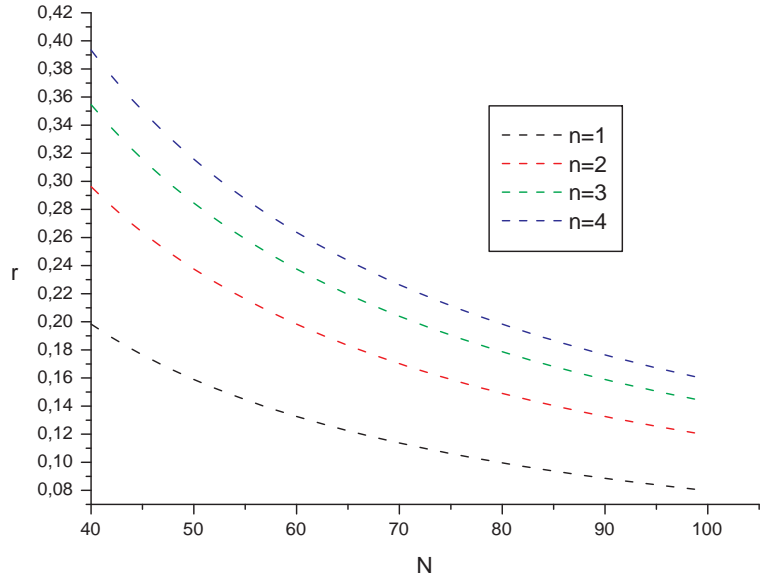}
\caption{$r$ versus $N$ for various values of $n.$We take $M\sim O(10^{15}$ $%
GeV),$ $\protect\lambda \sim O$ ($10^{68}GeV^{4})$ and $A\sim
10^{-13}M_{p}^{8}.$}
\label{fig2}
\end{figure}

In figure \ref{fig2} we have plotted $r$ as a function of e-folding number $%
N $ for different values of $n$. Generally, the ratio $r$ has a decreasing
behaviour with respect to $N$. We remarque also that, in order to confront $%
r $ with planck data we must have a large values of $N$ for the four values
of $n$.


\begin{figure}[]
\centering
\includegraphics[width=0.7\linewidth]{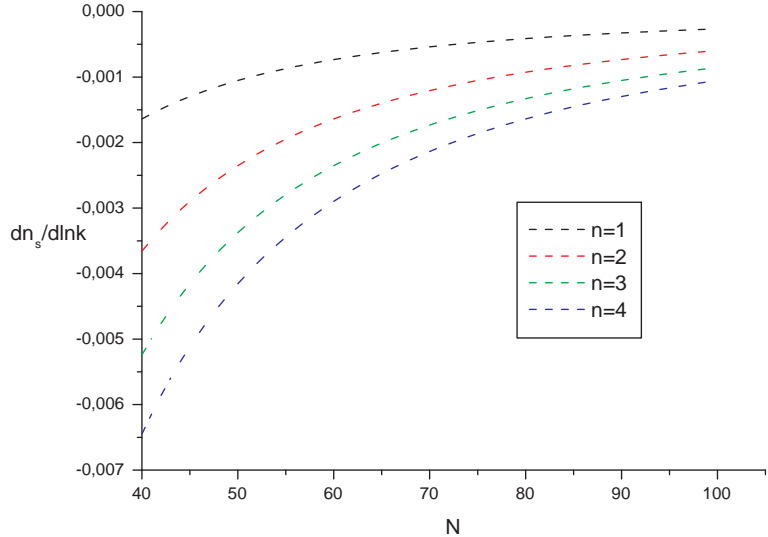}
\caption{$\frac{dn_{s}}{d\ln (k)}$ versus $N$ for various values of $n.$We
take $M\sim O(10^{15}$ $GeV),$ $\protect\lambda \sim O$ ($10^{68}GeV^{4})$
and $A\sim 10^{-13}M_{p}^{8}.$}
\label{fig3}
\end{figure}

Figure \ref{fig3} shows that the observable $\frac{dn_{s}}{d\ln (k)}$ is a
increasing function with respect to $N$. We have obtained extremely weak
values which are consistent with Planck data and that it gets smaller as we
increase the values of $n$.


To summarize this subsection, we have found that the results reviewed in the
context of the original chaplygin gas model are compatible with the latest
observational measurements for a particular choice of e-folding number $N$
and constant values of $n$.

In the following, we will study the case of generalized Chaplygin gas with $%
\alpha \neq 1$. We will discuss the effect of introducing the constant $%
\alpha $ on the peturbation spectrum of the model. Our results will be
compared to observations and we will show that the inflation can occur
successfully in relation to recent observations.

\subsection{ Generalized Chaplygin gas}

It is well known that generalized Chaplygin gas is one of the most natural
candidates dark energy models to explain the accelerated expansion of the
universe. In this context, these models have been extensively studied in the
literature. Morever, in the work \cite{BikashRDinda} the authors have
studied an inflationary scenario in the presence of Generalized Chaplygin
Gas in the light of the Planck and BICEP2 experiments, and have obtained the
constraints on the $n_{s}$ and $r$. In other work \cite{VFayaz} the
authors have examined the effect of anisotropy on generalized Chaplygin gas
scalar field and its interaction with other dark energy models, they
concluded that the increase in anisotropy leads to more correspondence
between the dark energy scalar field model and observational data. In the
other hand R. Herrera et all have considered an intermediate inflationary
universe model in the context of a generalized Chaplygin gas in the
slow-roll approximation, and have shown that the intermediate generalized
Chaplygin gas inflationary models are less restricted than analogous ones
standard intermediate inflationary models due to the introduction of $\alpha 
$ and $\beta $ parameters \cite{RamonHerrera}. In the context of
brane-inflationary background , the generalized cosmic Chaplygin model was
studied by A. Jawad et all \cite{AJawad} in the presence of chaotic
potential in the high-energy limit, various inflationary parameters was
evaluated and compared by planck data.

In this part, we study the generalized Chaplygin gas model in the presence
of monomial potential, using the basic formalism obtained in the section 2
in the context of high-energy limit. In this case, the scalar spectral index 
$n_{s}$, the ratio $r$ and the running of the scalar index $\frac{dn_{s}}{%
d\ln (k)}$ becomes:%
\begin{equation}
n_{s}=\frac{M_{p}^{2}\lambda }{2\pi (A+(M\phi _{\ast }^{n})^{\alpha +1})^{%
\frac{2}{\alpha +1}}}\left( -\frac{3n^{2}M^{3}\phi _{\ast }^{3n-2}}{%
(A+(M\phi _{\ast }^{n})^{\alpha +1})^{\frac{1}{\alpha +1}}}+n(n-1)M\phi
_{\ast }^{n-2}\right) +1
\end{equation}%
The ratio $r$ of tensor to scalar will be given by \ 
\begin{equation}
r=\frac{6M_{p}^{2}\lambda M^{2}n^{2}\phi _{\ast }^{2n-2}}{\pi (A+(M\phi
_{\ast }^{n})^{\alpha +1})^{\frac{3}{\alpha +1}}}
\end{equation}%
The running of the scalar spectral index will be in the following form 
\begin{equation}
\frac{dn_{s}}{d\ln (k)}=-\frac{M_{p}^{4}\lambda ^{2}n^{2}(2n+1)(n+2)}{8\pi
^{2}\phi _{\ast }^{4}(A+(M\phi _{\ast }^{n})^{\alpha +1})^{\frac{2}{\alpha +1%
}}}
\end{equation}%
We note that, as in the previous case, the inflaton value before the end of
inflation $\phi _{\ast }$, can be obtained numerically from Eq (\ref{NN}).

Based on the above formulas and to confront simultaniously the observables $%
n_{s},r$, and $\frac{dn_{s}}{d\ln (k)}$ with observations, we study the
relative variation of these parameters. We can see that these observables
depends on several parameters. Therefore, as the previous case we\ take\
the\ inflationary\ scale\emph{\ }$M\sim O(10^{15}$ $GeV),$ the brane tension
value $\lambda \sim O$ ($10^{68}GeV^{4})$ and $A\sim 10^{-13}M_{p}^{8}$ \cite{Herrera1}$,$ and $n=2$ which corresponds to chaotic case.

We will discuss some values of the above inflationary parameters in relation
with the e-folds number $N$ and the parameter $\alpha $. The central region,
given by the Planck data, of the spectral index $n_{s}$ $\in $ $\left[
0.959;0.969\right] $ gives $45<N<75$ for the values of $\alpha $\ in order
of $\alpha \sim O$ $(10^{-3})-$\emph{\ }$O$ $(10^{-1})$\emph{.} Regarding
the ratio of scalar to tensor curvature perturbation, the Planck constraint $%
r<0.11$, which requires large values of the folding number, that is, $N>80$,
and for $\frac{dn_{s}}{d\ln (k)}$ $\in \left[ -0.0166;-0.0039\right] $, we
get $56<N<150$ for large domain of value of $\alpha $ between \ $0<\alpha <1$%
. This shows that parameter $\alpha $ has a small influence on the
inflationary observables compared to the original Chaplygin gas. But we can
have the normal value which is most commonly used for the e-folding number,
required for solving the horizon and flatness problems.

\begin{figure}[]
\centering
\includegraphics[width=0.7\linewidth]{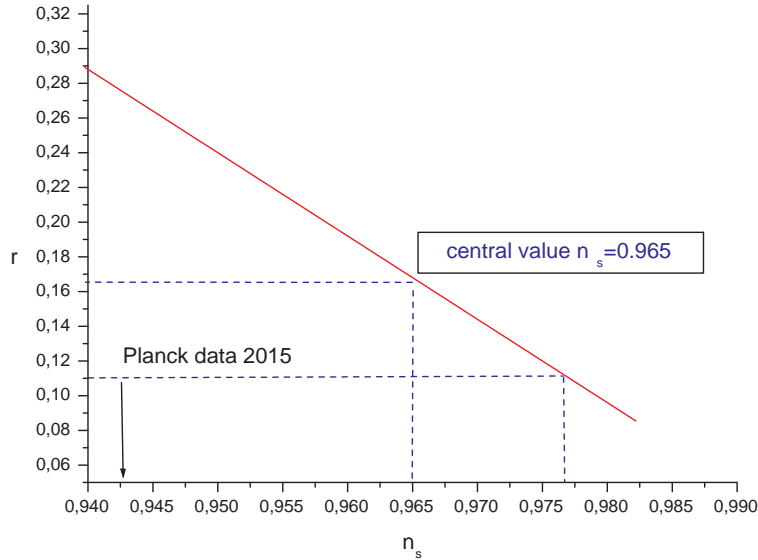}
\caption{Evolution of $r$ versus $n_{s}$ for various values of $N$. We take $%
M\sim O(10^{15}$ $GeV),$ $\protect\lambda \sim O$ ($10^{68}GeV^{4})$, $A\sim
10^{-13}M_{p}^{8}$ and $n=2$}
\label{fig4}
\end{figure}

Figure \ref{fig4} shows the variation of $n_{s}$ with respect to ratio $r$ ,
we remark that $n_{s}$ is an dicreasing linear function with $r.$ The ratio $%
r$ is compatible with Planck data where $r<0.11$ corresponds to $N>80$ and $%
n_{s}\succeq $ $0.975$ . The central value $n_{s}\simeq $ $0.965,$ where $%
N\sim 65$ corresponds to the ratio $r\sim 0.17$.


\begin{figure}[]
\centering
\includegraphics[width=0.7\linewidth]{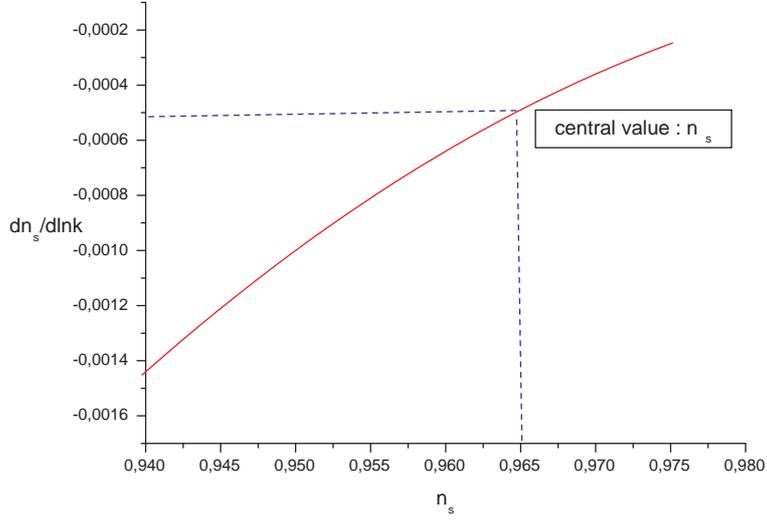}
\caption{ Evolution of $\frac{dn_{s}}{d\ln (k)}$ versus $n_{s}$ for various
values of $N$. We take $M\sim O(10^{15}$ $GeV),$ $\protect\lambda \sim O$ ($%
10^{68}GeV^{4})$, $A\sim 10^{-13}M_{p}^{8}$ and $n=2$}
\label{fig5}
\end{figure}

Figure \ref{fig5} presents\ the running of the scalar spectral index $\frac{
dn_{s}}{d\ln (k)}$ as function of $n_{s}.$ We see that the $\frac{dn_{s}}{
d\ln (k)}$ increases with respect to $n_{s},$ The central value of the
scalar spectral index $n_{s}\simeq $ $0.965$ corresponds to $N=65$, gives $%
\frac{dn_{s}}{d\ln (k)}\simeq -0.0005$, which is consistent with Planck data.


\begin{figure}[]
\centering
\includegraphics[width=0.7\linewidth]{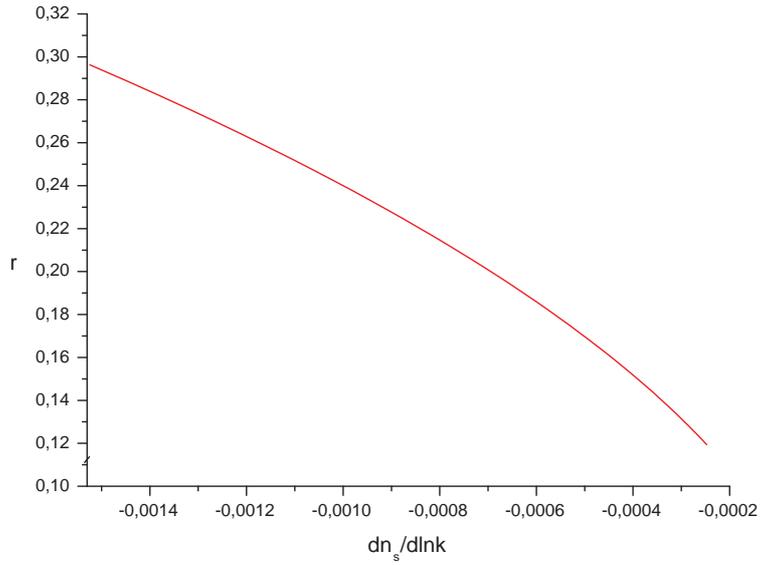}
\caption{Evolution of $\frac{dn_{s}}{d\ln (k)}$ versus $r$ for various
values of $N$. We take $M\sim O(10^{15}$ $GeV),$ $\protect\lambda \sim O$ ($%
10^{68}GeV^{4})$, $A\sim 10^{-13}M_{p}^{8}$ and $n=2$}
\label{fig6}
\end{figure}

In figure \ref{fig6} we have plotted the ratio $r$ according to the running
of the scalar spectral index $\frac{dn_{s}}{d\ln (k)},$ which is a
decreasing function with the variation of $N.$ For the ratio
tensor-to-scalar given by Planck corresponds to $N>80$ which gives a domain
of the running $\frac{dn_{s}}{ d\ln (k)}\preceq -0.00024$.

To summarize this subsection, our numerical calculation shows that, for a
particular choice of the parameters of the model, the obtained results are
compatibles with the Planck data, particularly for a suitable choice of the
e-folds number $N$. Also, the introduction of parametr $\alpha $ has a small
influence on the inflationary parameters.

\section{Conclusion}

In this paper, we have examined the Chaplygin gas model as a candidate for
inflation, which showed some important properties. We have considered the
original and generalized Chaplygin gas model in the Randall-Sundrum type II
braneworld using an monomial potential in the high-energy limit to study the
behaviors of inflationary spectrum perturbation parameters. We have found
that the inflationary parameters depend on a parameters space of the model.%
\emph{\ }In the original Chaplygin gas case, $\alpha =1,$ we have shown that
for $37<N<60$, the central value of the spectral index is reproduced
especially for $n=1;2$ and the running of the spectral index $\frac{dn_{s}}{%
d\ln (k)}$ is consistent with observations for large interval of $N$. A
confrontation with recent Planck data shows that the best fit is achieved
for large values of $N$, in particular for the ratio $r$. The case of
generalized Chaplygin gas predicts also a desirable values of ($n_{s}$; $r$
; $\frac{dn_{s}}{d\ln (k)}$) with recent observational data. In particular,
the central value, $n_{s}\simeq $ $0.965$, where $N=65$ corresponds to the
value of $\alpha $ in order of $O$ $(10^{-3})-$\emph{\ }$O$ $(10^{-1})$ and
for the ratio $r$ is consistent with Planck 2015 $r<0.11$, is compatible in
the case where $N\succeq 80$. The values of the tensor to scalar ratio $r$
and running of the spectral index $\frac{dn_{s}}{d\ln (k)}$ are in excellent
agreement with the latest observations of the Planck satellite for a
particular choice of the parameter space of the model.

\bibliographystyle{unsrt}
\bibliography{biblio}

\begin{thebibliography}{10}

\bibitem{AH}
A.~Guth.
\newblock Inflationary universe: A possible solution to the horizon and
  flatness problems.
\newblock {\em Phys.Rev.}, D 23:347 -- 356, 1981.

\bibitem{AD}
A.~D. Linde.
\newblock {\em Particle Physics and Inflationary Cosmology}.
\newblock Harwood Academic, Switzerland, 5 edition, 1990.

\bibitem{Liddle}
Liddle~A. R. and Lyth~D. H.
\newblock {\em Cosmological Inflation and Large-Scale Structure}.
\newblock Cambridge University Press, Cambridge, 2000.

\bibitem{DN}
M.C. Bento, O.~Bertolami, and A.A. Sen.
\newblock Wmap constraints on the generalized chaplygin gas model.
\newblock {\em Phys.Lett.}, B575:172--180, 2003.

\bibitem{DM}
Rachel Bean and Dore Olivier.
\newblock Are chaplygin gases serious contenders for the dark energy?
\newblock {\em Phys. Rev.}, D 68(2):023 -- 515, 2003.

\bibitem{Bouhmadi}
Mariam Bouhmadi-Lopez and Ruth Lazkoz.
\newblock Chaplygin dgp cosmologies.
\newblock {\em Phys. Lett.}, B 654:51--57, 2007.

\bibitem{OLuongo}
O.~Luongo and H.~Quevedo.
\newblock A unified dark energy model from a vanishing speed of sound with
  emergent cosmological constant.
\newblock {\em Int. J. Mod. Phys.}, D 23:1450012, 2014.

\bibitem{LXu}
L.~Xu, Y.~Wang, and H.~Noh.
\newblock Modified chaplygin gas as a unified dark matter and dark energy model
  and cosmic constraints.
\newblock {\em Eur. Phys. J.}, C 72:1931, 2012.

\bibitem{MubasherJamil}
Mubasher Jamil.
\newblock Interacting new generalized chaplygin gas.
\newblock {\em Int. J. Theor. Phys.}, 49:62--71, 2010.

\bibitem{JGleyzes}
J.~Gleyzes, D.~Langlois, and F.~Vernizzi.
\newblock Interacting new generalized chaplygin gas.
\newblock {\em Int. J. Mod. Phys.}, D 23:1443010, 2014.

\bibitem{EGuendelman}
E.~Guendelman, E.~Nissimov, and S.~Pacheva.
\newblock Unified dark energy and dust dark matter dual to quadratic purely
  kinetic k-essence.
\newblock {\em Eur. Phys. Jour.}, C 76:90, 2016.

\bibitem{Orlando}
O.~Luongo and H.~Quevedo.
\newblock An expanding universe with constant pressure and no cosmological
  constant.
\newblock {\em Astroph. sp. sci.}, 338(2):345--349, 2012.

\bibitem{MBronstein}
M.~Bronstein.
\newblock {\em Phys. Zeit}.
\newblock Sowejt Union 3, 5 edition, 1993.

\bibitem{MRSetare}
M.~R. Setare, J.~Sadeghi, and A.~R. Amani.
\newblock Title.
\newblock {\em Phys. Lett.}, B673, 2009.

\bibitem{PFGonzalez}
P.~F. Gonzalez-Diaz.
\newblock Title.
\newblock {\em Phys. Rev.}, D 62:023513, 2000.

\bibitem{EOKahya}
E.O. Kahya and Pourhassan B.
\newblock The universe dominated by the extended chaplygin gas.
\newblock {\em Modern Physics Letters A}, 30(13):1550070, 2015.

\bibitem{FRW}
B.~Pourhassan and E.~O. Kahya.
\newblock Frw cosmology with the extended chaplygin gas.
\newblock {\em Advances in High Energy Physics}, 2014.

\bibitem{Benaoum}
H.~B. Benaoum.
\newblock Modified chaplygin gas cosmology with bulk viscosity gas.
\newblock {\em International Journal of Modern Physics}, D 23(10):1450082,
  2014.

\bibitem{RZarrouki}
R.~Zarrouki and M.~Bennai.
\newblock Chaplygin gas braneworld inflation according to wmap7.
\newblock {\em Phys. Rev.}, D82:123506, 2010.

\bibitem{JRVillanueva}
R.~Zarrouki and M.~Bennai.
\newblock The generalized chaplygin--jacobi gas.
\newblock {\em JCAP}, 07:045, 2015.

\bibitem{SBNassur}
S.~B. Nassur, M.~J.~S. Houndjo, I.~G. Salako, and J.~Tossa.
\newblock Interactions of some fluids with dark energy in f (t) theory.
\newblock 2016.
\newblock arXiv preprint arXiv:1601.04538.

\bibitem{SouravDutta}
S.~Dutta, S.~Mukerji, and S.~Chakraborty.
\newblock An attempt for an emergent scenario with modified chaplygin gas.
\newblock {\em Advances in High Energy Physics}, 2016.

\bibitem{BPourhassan}
B.~Pourhassan and E.~O. Kahya.
\newblock Frw cosmology with the extended chaplygin gas.
\newblock {\em Advances in High Energy Physics}, 2014.

\bibitem{OBertolami}
O.~Bertolami and V.~Duvvuri.
\newblock Chaplygin inspired inflation.
\newblock {\em Phys. Lett.}, B 640:121--125, 2006.

\bibitem{Herrera1}
R.~Herrera.
\newblock Chaplygin inflation on the brane.
\newblock {\em Phys. Lett.}, B 664:149--153, 2008.

\bibitem{Herrera2}
R.~Herrera.
\newblock Tachyon-chaplygin inflation on the brane.
\newblock {\em General Relativity and Gravitation}, 41(6):1259--1271, 2009.

\bibitem{Bento}
M.~C. Bento, O.~Bertolami, and A.~A. Sen.
\newblock Generalized chaplygin gas, accelerated expansion, and
  dark-energy-matter unification.
\newblock {\em Phys. Rev.}, D 66(4):043507, 2002.

\bibitem{Planck2015}
Planck Collaboration.
\newblock Planck 2015 results. xiii. cosmological parameters.
\newblock 2015.
\newblock arXiv:1502.01589v2 [astro-ph.CO].

\bibitem{David}
D.~H. Lyth and A.~Riotto.
\newblock Particle physics models of inflation and the cosmological density
  perturbation.
\newblock {\em Physics Reports}, 314(1):1--146, 1999.

\bibitem{Maartens}
R.~Maartens.
\newblock Brane-world gravity.
\newblock {\em Living Rev. Rel.}, 7, 2004.

\bibitem{Langlois}
D.~Langlois, R.~Maartens, and D~Wands.
\newblock Gravitational waves from inflation on the brane.
\newblock {\em Phys. Lett.}, B 489:259--267, 2000.

\bibitem{Braneworld}
P.~Brax, C.~Bruck, and A.~Davis.
\newblock Brane world cosmology.
\newblock {\em Rept. Prog. Phys.}, 67:2183--2232, 2004.

\bibitem{PingXi}
Xi~Ping and Li~Ping.
\newblock Reexamining generalized chaplygin gas with the sign-changeable
  interaction.
\newblock 2015.
\newblock arXiv:1510.02859v1 [gr-qc].

\bibitem{RSII1}
L.~Randall and R.~Sundrum.
\newblock A large mass hierarchy from a small extra dimension.
\newblock {\em Phys. Rev. Lett.}, 83:3370--3373, 1999.

\bibitem{P}
P.~Binetruy, C.~Deffayet, U.~Ellwanger, and D.~Langlois.
\newblock An alternative to compactification.
\newblock {\em Phys. Lett.}, B477:285, 2000.

\bibitem{JCFabris}
J.~C. Fabris.
\newblock 2001.
\newblock arXiv:gr-qc/0103083v1.

\bibitem{HongshengZhang}
Hongsheng Zhang.
\newblock 2006.
\newblock arXiv:astro-ph/0509895v2.

\bibitem{MayraJReyes}
Mayra~J. Reyes-Ibarra.
\newblock In {\em AIP Conference Proceedings}, 2010.
\newblock doi: 10.1063/1.3473869.

\bibitem{Zarrouki}
R.~Zarrouki, Z.~Sakhi, and M.~Bennai.
\newblock On braneworld inflation models in light of wmap7 data.
\newblock {\em General Relativity and Gravitation}, 43(5):1515--1528, 2011.

\bibitem{BikashRDinda}
B.~R. Dinda, S.~Kumar, and A.~A. Sen.
\newblock Inflationary generalized chaplygin gas and dark energy in light of
  the planck and bicep2 experiments.
\newblock {\em Phys. Rev.}, D 90(8):083515, 2014.

\bibitem{VFayaz}
V.~Fayaz, H.~Hossienkhani, and A.~Jafari.
\newblock Effectof anisotropy on the generalized chaplygin gas scalar field and
  its interaction with other dark energy models.
\newblock {\em The European Physical Journal Plus}, 132(4):193, 2017.

\bibitem{RamonHerrera}
R.~Herrera, M.~Olivares, and N.~Videla.
\newblock Intermediate-generalized chaplygin gas inflationary universe model.
\newblock {\em The European Physical Journal C}, 73(1):2295, 2013.

\bibitem{AJawad}
A.~Jawad, S.~Rani, and S.~Mohsaneen.
\newblock Generalized cosmic chaplygin inflationary model on the brane.
\newblock {\em The European Physical Journal Plus}, 131(7):1--7, 2016.

\end{thebibliography}

\textbf{We declares that there is no conflict of interest regarding the
publication of this paper}
\end{document}